\documentclass[12pt]{article}

\usepackage{amsmath}
\usepackage{amssymb}

\usepackage[T1,T2A]{fontenc}
\usepackage[utf8]{inputenc}
\usepackage[english]{babel}
\usepackage{color}
\usepackage{graphicx}

\usepackage{fullpage}

\def\M2Pl{M^2_{\mathrm{Pl}}}
\def\MPl{M_{\mathrm{Pl}}}

\def\L{{\cal {\bar L}}}

\def\d{\partial}
\def\D{\nabla}
\def\Y{\Upsilon}
\def\ls{\lambda_s}

\def\bg{{\bar g}}
\def\bR{{\bar R}}

\begin{document}

\title{Multiscalar-metric gravity: merging gravity,\\ dark energy and dark matter}

\author{Yu.\ F.\ Pirogov\footnote{Yury.Pirogov@ihep.ru}\, and\, 
        O.\ V.\ Zenin\footnote{Oleg.Zenin@ihep.ru}\\
		{\it NRC ``Kurchatov Institute'' -- IHEP, Protvino, Russia}}

\date{}
\maketitle

\begin{center}
	{\it
	 Reported at the 22nd Lomonosov Conference on Elementary Particle Physics,\\
	 Moscow, Russia, August 21-27, 2025
	}
\end{center}

\begin{abstract}
The status of a modification of General Relativity (GR) -- Spontaneously Broken Relativity (SBR) 
-- for merging gravity, dark energy (DE) and dark matter (DM) is presented.
The modification is principally grounded on a multiscalar-metric concept of spacetime
endowed with two dynamical structures: a basic metric and a set of the reversible multiscalar fields.
The latter ones serve geometrically as exceptional dynamical coordinates among arbitrary kinematical/observer's ones
and physically as a kind of gravitational Higgs fields producing possible spontaneous breaking of the symmetry (SSB) of relativity.
The effective field theory (EFT) of the extended gravity based on the multiscalar-metric spacetime  -- 
the metagravity -- is discussed and some of its particular realizations beyond GR as SBR are explicated.
Physically, SBR results in appearance of massive tensor and scalar gravitons.
Some of the emerging consequences, problems and prospects for SBR 
for future deeper unification of gravity with matter in the context of the relativity and internal symmetries breaking are shortly discussed.
\end{abstract}

\begin{center}
 {\small Keywords: 
  multiscalar-metric spacetime; metagravity; spontaneously broken relativity;\\
  dark energy; dark matter; scalar graviton
 }
\end{center}

\maketitle

\section{Introduction: dark texture vs observed  structures}

The existence of (cold) DM is strongly motivated by cosmological and astrophysical observational data analyzed 
in the framework of General Relativity (GR) and the Standard Model of particle physics (SM)~\cite{PDG2024}.
Nevertheless, the DM phenomenon raises a number of theoretical and experimental issues.
The SM contains no satisfactory DM candidate, and there is no obvious path of SM extension providing it.
Experimental searches for DM particles did not yield any positive signal so far.
Commonly accepted cold DM, irrespective of its particular nature, exhibits dynamical problems at scales $\le$1~Mpc~\cite{PDG2024}. 

Given the DM problems emerging in the framework of GR and (extended) SM, one may naturally search the alternative as dark degrees of freedom in a modified gravity.
GR, as an effective field theory (EFT) with the gauge symmetry group of diffeomorphisms (Diff), leaves only two on-shell degrees of freedom for the massless tensor graviton.
Additional gravitational degrees of freedom may originate from a spontaneous breaking of the Diff gauge symmetry, 
with extra polarizations of now massive tensor graviton and a physical scalar gravi-Higgs boson (scalar graviton) serving as dark energy (DE) and (part of) DM candidates.
To this end, a modified concept of a tagged spacetime manifold is considered as a dark texture supporting the observed matter structures~\cite{Pirogov2015,Pirogov2017,Pirogov2018}.

\section{Multiscalar-metric/tagged spacetime}

The generally adopted approach to gravity in the framework of GR and its immediate modifications 
is based on the concept of a topological manifold endowed with piecewise arbitrary kinematical/observer's 
coordinates $x^\mu$ ($\mu = 0,1,\dots d-1$; $d$ is the spacetime dimension)
and a dynamical metric $g_{\mu\nu}(x)$ with $g \equiv \mathrm{det}g_{\mu\nu} \ne 0$.
To go beyond GR with merging gravity and dark components,
an approach was proposed
based on the concept of a modified spacetime tagged additionally with some exceptional dynamical coordinates $z^\alpha$ ($\alpha = 0,1,\dots d-1$)
among the arbitrary kinematical ones $x^\mu$.
The exceptional coordinates $z^\alpha$ are introduced piecewise through a set of reversible multiscalar fields $Z^a$ ($a = 0,1,\dots d-1$) with dimension of length,
$z^{\alpha}(x) \equiv \delta^\alpha_a Z^a(x)$.
To ensure the homogeneity, $Z^a$ are supposed to enter only through derivatives with $\mathrm{det}(\partial_\mu Z^a) \ne 0$.
To preserve the geometrical interpretation, they are further supposed to enter through 
the auxiliary metric 
$\zeta_{\mu\nu} \equiv \partial_\mu Z^a \partial_\nu Z^b \eta_{ab}$ ($\zeta \equiv \mathrm{det}\zeta_{\mu\nu} \ne 0$),
where $\eta_{ab}$ is the Minkowski symbol corresponding to the auxiliary global (piecewise) Lorentz group.

\section{Effective field theory of metagravity}

The phenomenon of merging gravity and dark components based on the multiscalar-metric manifold may be called {\it metagravity}.
The pure metagravity is adopted to be described in observer's arbitrary kinematic coordinates $x^\alpha$ by an EFT depending on the basic dynamical metric $g_{\mu\nu}(x^\alpha)$ 
and the auxiliary metric $\zeta_{\mu\nu}(x^\alpha)$ through the action 
$S[g_{\mu\nu}, \zeta_{\mu\nu}] = \int{\cal L}d^dx$, where ${\cal L}$ is Lagrangian density.
In the low energy approximation, $\zeta_{\mu\nu}$ is taken to enter observables only through the correlator with the basic metric:
\begin{equation} \label{eq:correlator}
	{\ae}^\mu_\nu \equiv g^{\mu\lambda} \zeta_{\lambda\nu} \, .
\end{equation}
%As a simplest possibility, an explicit dependence on $\zeta_{\mu\nu}$ is assumed to be absent.
%
More particularly, the dependence on the correlator may be expressed through its matrix logarithm,
\begin{equation} \label{eq:log-correlator}
	{\AE}^\mu_\nu \equiv \log {\ae}^{\mu}_\nu \, .
\end{equation}
The latter can be decomposed through its trace and the traceless part,
\begin{equation}\label{eq:tilde-AE}
	{\tilde{\AE}}^{\mu}_{\nu} \equiv {\AE}^\mu_\nu - \frac{1}{d} \delta^\mu_\nu \mathrm{Tr} {\AE}^\lambda_\rho \, .
\end{equation}
The trace transforms as a scalar due to 
\begin{equation}
	\mathrm{Tr} {\AE}^\mu_\nu = \log \mathrm{det}\left({\ae}^\mu_\nu\right) = \log(\zeta/g) \, .
\end{equation}
Hence, one can choose as an independent effective variable the scalar field,
\begin{equation} \label{eq:def-sigma}
	\sigma \equiv \frac{1}{2}\log(g/\zeta) \, , 
\end{equation}
which may be called the {\it scalar graviton},
whereas the traceless part (\ref{eq:tilde-AE}) may be associated with the dark energy.
Note that under (patchwise) constant rescaling $Z^a \to e^{\gamma} Z^a$, the scalar graviton shifts as $\sigma \to \sigma - \gamma d$, whereas ${\tilde{\AE}}^{\mu}_{\nu}$ remains invariant.
Assuming the emerging shift symmetry  as an approximate dynamical one implies that $\sigma$ should enter the Lagrangian predominantly through derivatives,
while the DE component ${\tilde{\AE}}^{\mu}_{\nu}$ is not suppressed to enter directly.

The pure metagravity is self-sufficient, but in reality it should be supplemented by conventional matter (CM) and hypothetical DM fields,
in particular, serving as a source of the scalar graviton.
The mixed field-theoretic--particle approach conceivably provides the inevitable route to merging gravity with dark components.

For definiteness, we restrict consideration to dimension $d=4$.

\section{Spontaneously broken relativity (SBR)}

\subsection{Minimal realization}

The effective Lagrangian density of pure metagravity looks like
\begin{equation} \label{eq:action}
	{\cal L} =  \L(\bar{g}_{\mu\nu}, \tilde{\bar{\AE}}^\mu_\nu, \sigma)\, \sqrt{-\bar{g}}\,  ,
\end{equation}
where $\L$ is the Lagrangian depending on 
the effective (i.e. the one defining the observables) metric $\bg_{\mu\nu}$  chosen as
\begin{equation} \label{eq:def-gbar}
 \bar{g}_{\mu\nu} =  e^{\bar\gamma} g_{\mu\nu}\,\, ({\bar g}^{\mu\nu} \equiv {\bar g}^{-1\, \mu\nu}) \, . 
\end{equation} 
For simplicity, the gravity form-factor defining the type of the effective metric is chosen as $\bar \gamma = \bar \gamma(\sigma)$ ($\bar \gamma(0) = 0$), 
with $\sigma$ generally depending on matter.
The two marginal cases are of special interest:
at $\bar \gamma \equiv 0$ the effective metric presents the GR metric $\bg_{\mu\nu}\equiv g_{\mu\nu}$,
and at $\bar \gamma \equiv -\sigma/2$ the metric of Weyl-transverse gravity, $\bg_{\mu\nu} \equiv (g_{\mu\nu} / (-g)^{1/4}) (-\zeta)^{1/4}$. 

It is convenient to define the effective tensor variables,
\begin{equation}\label{eq:def-hbar}
	\bar{\ae}^\mu_\nu \equiv \bar{g}^{\mu\lambda} \zeta_{\lambda\nu} \, , \,\, %
	\bar{\AE}^\mu_\nu \equiv \log \bar{\ae}^\mu_\nu \, .
\end{equation}
The traceless part of $\bar{\AE}^\mu_\nu$ coincides with (\ref{eq:tilde-AE}) due to $Z^a$ scale invariance of the latter:
$\tilde{\bar{\AE}}^\mu_\nu = \tilde{\AE}^\mu_\nu$.
The EFT of metagravity {\it a priori} admits multiple realizations. 
Somewhat restricted but sufficiently general ones are considered in what follows.

For consistency, the gravity theory
is considered as a gauge one corresponding to a spontaneosly broken relativity symmetry, with $Z^a$ playing the r\^ole of Higgs fields for gravity.
Explicit form of gauge transformations for the basic and effective variables can be found in Ref.~\cite{Pirogov2019}.
From particle physics viewpoint, 3 combinations of $Z^{a}$ components become additional components of a massive tensor graviton, while 
the remaining one combination of $Z^a$ may be considered as the explicit physical Higgs field for gravity.
For definiteness, $\tilde{\bar{\AE}}^\mu_\nu$ and $\sigma$ are supposed not to be connected with CM and directly coupled only to tensor gravity and DM.

The minimal Lagrangian is chosen as
 \begin{eqnarray}
	 \L_{\mathrm{min}}      &=& \L_G +  \L_{SG} + \L_{DE} + \L_{CM} + \L_{DM} \label{eq:L} \, ,  \\
	 \L_G    &=& -\frac{1}{2} \M2Pl\, \bar R(\bar g_{\mu\nu}) \label{eq:LG} \, ,  \\
	 \L_{SG} &=& \frac{\M2Pl}{2} \Upsilon^2 \bar g^{\mu\nu} \partial_\mu \sigma \partial_\nu \sigma \label{eq:LSG} \, , \\ 
	 \L_{DE} &=& - \bar V_{DE} (\tilde{\bar{\AE}}^\mu_\nu, \sigma) = %\nonumber \\
	         %&=& %
			 - \M2Pl \left[ \bar v_{{\ae}}({\tilde{\bar{\AE}}}^\mu_\nu) + v_\sigma(\sigma) \right] \label{eq:LDE}  , \,\,\,\,\,\,\,  \\
	 \L_{CM} &=& \L_{CM}(\bar g_{\mu\nu}, \phi_{CM}) \label{eq:LCM} \, ,   \\
	 \L_{DM} &=& \L_{DM/CM}(\bg_{\mu\nu}, \phi_{DM}, \phi_{CM}) -  \d_\mu\sigma J^\mu_{DM} \label{eq:LDM} \, . \,\,\,\,\,\,\,\,
\end{eqnarray}
Here $\MPl = (8\pi G_N)^{-1/2} = 2.4 \times 10^{18}$~GeV is the reduced Planck mass.
$\bR_{\mu\nu}$ in (\ref{eq:LG}) is Riemann curvature tensor expressed via the effective metric.
The dimensionless constant $\Upsilon$ in (\ref{eq:LSG}) characterizes the coupling between scalar and tensor gravity and may be expressed as $\Upsilon = M_S / \MPl$, where  $M_S$ is some scalar gravity scale.
Derivatives of ${\tilde{\bar{\AE}}}^\mu_\nu$ are omitted for simplicity.
The dark energy potential %$\bar V_{DE}(\tilde{\bar{\AE}}^\mu_\nu, \sigma)$ 
(\ref{eq:LDE}) responsible for the spontaneous relativity breaking~\cite{Pirogov2021}
generally includes a constant term which is nothing but the cosmological constant (CC) $\M2Pl \bar \Lambda$, 
and a dynamical part depending on traces of products of $\tilde{\bar{\AE}}^{\mu}_{\nu}$ and the scalar graviton field.
For simplicity, we factorize $\bar V_{DE}$ into pure scalar graviton and pure DE terms. 
%\footnote{%
%Under such a division,  the shift symmetry $\sigma \to \sigma + const$ is violated only by $v_\sigma(\sigma)$.
%}
In what follows, we do not assume a particular form of the $\bar v_{{\ae}}$ and $v_\sigma$.
An example of $\bar V_{DE}$ for a consistent implementation of a spontaneosly broken Weyl-transverse relativity can be found in \cite{Pirogov2021}.
CM fields $\phi_{CM}$ and possible additional DM fields $\phi_{DM}$  enter, respectively, in (\ref{eq:LCM}) and (\ref{eq:LDM}).

\subsection{Field equations}
Extremizing the action with Lagrangian density~(\ref{eq:L}) with respect to the basic metric $g^{\mu\nu}$ 
gives Einstein's equations in terms of the effective metric $\bar g^{\mu\nu}$ and $\sigma$ (with traceless terms explicitly separated):
\begin{equation} \label{eq:Einstein-traceless}
%\begin{eqnarray} \label{eq:Einstein-traceless}
	\M2Pl \left( \bar R_{\mu\nu} - \frac{1}{4} \bar g_{\mu\nu} \bar R \right) %
	= %
	\bar T_{\mu\nu} - \frac{1}{4} \bar g_{\mu\nu} \bar T   %
	+ \, \bar g_{\mu\nu} %
	    \left[ %
     			\frac{1}{4} (1 + 2\bar \gamma') \left( \M2Pl \bar R + \bar T \right) - \frac{1}{\sqrt{-\bar g}} \frac{\delta \sqrt{-\bar g} \L}{\delta\sigma} %
		\right] \,\, , 
%\end{eqnarray}
\end{equation}
where
\begin{eqnarray}
	\L &=& \L_{CM} + \L_{DM} + \L_{SG} + \L_{DE} \, , \label{eq:Ltot}%
	   \\
	\bar T_{\mu\nu} &=& %
		  \frac{2}{\sqrt{-\bar g}} \frac{\delta \sqrt{-\bar g}\L}{\delta \bar g^{\mu\nu}} \, , \label{eq:T} %
		\\
	\bar \gamma' &\equiv& d\bar \gamma(\sigma)/d\sigma\, . \label{eq:wprime}
\end{eqnarray}
SG, DE and DM contributions to $\bar T_{\mu\nu}$ read, respectively:
\begin{eqnarray}
	\bar T_{SG\, \mu\nu} &=&   \M2Pl \Upsilon^2 %
									\left( %
									       \d_\mu\sigma \d_\nu\sigma %
	                                     - \bg_{\mu\nu} \frac{1}{2}\bg^{\lambda\rho} \d_\lambda\sigma \d_\rho\sigma %
									\right) %
									 , \,\,\,\,\,\,\,\,\, \\
	\bar T_{DE\, \mu\nu} &=&  \M2Pl %
								\left[ -2 \frac{\d\bar v_{\ae}}{\d{\bar{\ae}}^\mu_\beta} \zeta_{\nu\beta} %
									   + \bg_{\mu\nu} \left(\bar v_{\ae} + v_\sigma \right) %
								\right] %
								   \, , \\
	\bar T_{DM\, \mu\nu} &=& 	\frac{2}{\sqrt{-\bg}} \frac{\delta \sqrt{-\bg} \L_{CM/DM}}{\delta\bg^{\mu\nu}}  %
								 - \d_{(\mu}\sigma {J_{DM}}_{\nu)} %
						  	 -  \bg_{\mu\nu} (-\d_\lambda\sigma J^{\lambda}_{DM}) %
	 - \d_\lambda\sigma \frac{\delta {J_{DM}}_\rho}{\delta \bar g^{\mu\nu}} \bar g^{\lambda\rho} %
	  \, .
\end{eqnarray}

Extremizing the action with respect to $Z^a$ fields gives:\footnote{%
	In (\ref{eq:Z-diff}), (\ref{eq:sigma-wave}),
	$\bar v_{\ae}(\tilde{\bar{\AE}}^\mu_\nu)$ is varied by $\bar{\ae}^\mu_\nu$ rather than by $\tilde{\bar{\AE}}^\mu_\nu$ to obtain more compact expressions.
}
\begin{equation} \label{eq:Z-diff}
	\D_\lambda \left\{ %
		               \eta_{ab} \d_\rho Z^b %
					   \left[ %
						     {\zeta^{-1}}^{\lambda\rho} %
							  \left( %
							        \frac{\bar \gamma'}{2} \left( \M2Pl \bar R + \bar T \right) %
									- \frac{\delta \left(\sqrt{-\bar g} \L\right)}{{\sqrt{-\bar g}}\, \delta\sigma} %
							  \right) %
							- \M2Pl %
							\left( %
									  \frac{\d \bar v_{\ae}}{\d {\bar {\ae}}^\mu_\rho} \bar g^{\mu\lambda} % 
									+ \frac{\d \bar v_{\ae}}{\d {\bar {\ae}}^\mu_\lambda} \bar g^{\mu\rho} % 
							\right) %
					   \right] %
		       \right\} = 0 \, . 
\end{equation}
The equation (\ref{eq:Z-diff}) can be immediately integrated, so that 
with account of Eqs.~(\ref{eq:LSG})--(\ref{eq:LDE}) and (\ref{eq:Einstein-traceless})  
one obtains the equation for the scalar graviton field:
\begin{equation}\label{eq:sigma-wave}
	\frac{\M2Pl\Upsilon^2}{\sqrt{-\bar g}} \d_\mu %
		\left[ \sqrt{-\bar g} \bg^{\mu\nu} \d_\nu \sigma \right] %
		+ \frac{\d V^{\mathrm{eff}}_\sigma}{\d\sigma} %
		= \D_\mu J^\mu_{DM}  
	    + (1 + 2\bar \gamma') %
		  \M2Pl %
	      \left[ %
			  \frac{1}{2} \frac{\d \bar v_{\ae}}{\d {\bar {\ae}}^\lambda_\rho} {\bar {\ae}}^\lambda_\rho %
			  + \eta_{ab}\d_\lambda Z^a {f^b}^\lambda %
		 \right] \, ,  
\end{equation}
where ${f^b}^\lambda$ is an arbitrary bi-vector, such that $\D_\lambda {f^b}^\lambda = 0$ and ${f^b}^\lambda \d_\lambda Z^a = {f^a}^\lambda \d_\lambda Z^b$ ($a \ne b$),
and the effective potential of the scalar graviton is 
\begin{equation}\label{eq:Veff}
	V^{\mathrm{eff}}_\sigma(\sigma) = \M2Pl \left[ v_\sigma(\sigma) + \ls e^{-2\bar \gamma - \sigma} \right]\, .
\end{equation}
$\ls$ in (\ref{eq:Veff}) is an arbitrary integration constant. 
Note that for $\bar \gamma \ne -\sigma/2$ the presence of the integration constant $\ls \ne 0$ makes the scalar graviton 
self-interacting even if its Lagrangian potential $\M2Pl v_\sigma(\sigma) \equiv 0$.

In what follows, two representative cases related to the choice of $\bar \gamma$ are considered.

\paragraph*{Exceptional case: a spontaneosly broken Weyl-transverse relativity (SBWTR), $\bar \gamma = -\sigma/2$,}
which represents a generalization of the unimodular gravity.
In this special case the structure of the theory is greatly simplified~\cite{Pirogov2021}.
Einstein's equation~(\ref{eq:Einstein-traceless}) takes the form:
\begin{equation}\label{eq:Einstein-Weyl}
	\M2Pl \left(\bar R_{\mu\nu} - \frac{1}{2} \bar g_{\mu\nu} \bar R \right) %
	= \bar T_{\mu\nu} - \frac{1}{4} \bar g_{\mu\nu} \left(\M2Pl \bar R + \bar T\right) \, ,
\end{equation}
where $\bar T_{\mu\nu}$ is defined by (\ref{eq:T}).
The Lagrangian CC term $\M2Pl \bar \Lambda$ does not contribute to the effective energy-momentum tensor in the r.h.s. of Eq.~(\ref{eq:Einstein-Weyl}),
which reflects the invariance of the measure $\sqrt{-\bar g} d^4x$ under Diff transformations.\footnote{%
The effective metric is  $\bar g_{\mu\nu} \equiv e^{-\sigma/2} g_{\mu\nu} \equiv \left( g_{\mu\nu} / (-g)^{1/4} \right) (-\zeta)^{1/4}$, 
so that $\sqrt{-\bar g} \equiv \sqrt{-\zeta}$ and the Lagrangian CC term 
$ \sqrt{-\bar g} \M2Pl \bar\Lambda \equiv \sqrt{-\zeta} \M2Pl \bar\Lambda$ does not vary with $g_{\mu\nu}$.
}	
Imposing an additional condition, e.g., the covariant conservation of the energy-momentum tensor $\bar T_{\mu\nu}$,
one obtains the constraint:
\begin{equation}\label{eq:Weyl-CC}
	\d_\mu \left(\M2Pl \bar R + \bar T \right) = 0\, , 
\end{equation}
bringing back an effective CC into r.h.s. of Eq.~(\ref{eq:Einstein-Weyl}), now as an arbitrary integration constant unrelated to the Lagrangian CC,
which may open a possible path to explain the CC issues \cite{Pirogov2021}.
The massive tensor graviton appears via SSB mechanism. 
A particular ghost-free SBWTR realisation with a smooth $m_g \to 0$ limit was considered in Ref.~\cite{Pirogov2021}.

\paragraph*{Conventional case: spontaneously broken GR (SBGR), $\bar \gamma = 0$,  $\bar g_{\mu\nu} \equiv g_{\mu\nu}$.}
In SBGR,  Einstein's equations~(\ref{eq:Einstein-traceless}) take a GR-like form:
\begin{equation}\label{eq:Einstein-GR}
	\M2Pl \left(\bar R_{\mu\nu} - \frac{1}{2} \bar g_{\mu\nu} \bar R \right) %
	= \bar T_{\mu\nu} - \bar g_{\mu\nu} \frac{1}{\sqrt{-\bar g}} \frac{\delta \sqrt{-\bar g} \L}{\delta\sigma} \, , 
\end{equation}
with the energy--momentum tensor given by (\ref{eq:T}). 
The equation for the scalar graviton is given by (\ref{eq:sigma-wave}) with $\bar \gamma = 0$.

\subsection{SBGR: scalar graviton dark holes}

To illustrate the discussed theory, a local stationary spherically symmetric solution is considered in the SBGR context ($\bg_{\mu\nu} \equiv g_{\mu\nu}$).
To study static spherically symmetric vacuum solutions of Eqs.~\ (\ref{eq:Einstein-GR}), (\ref{eq:sigma-wave})
merging a GR-like black hole (BH) and a scalar graviton halo~\cite{Pirogov2012},
we define the line element 
$ds^2 \equiv g_{\mu\nu}\, dx^\mu dx^\nu$
in polar coordinates $r, \theta, \phi$ in the reciprocal gauge ($AB = 1$):
\begin{equation}
	ds^2 = A(r) dt^2 - C(r) r^2 (\sin^2 \theta\, d\phi^2 + d\theta^2) %
	   - A^{-1}(r) dr^2 \, .
\end{equation}
For simplicity, Lagrangian potential terms for DE and the scalar graviton (\ref{eq:LDE}) are omitted,
which, particularly, implies setting the CC to zero.
We also neglect possible DM sources of the scalar graviton: $J^\mu_{DM} = 0$. 
In (\ref{eq:sigma-wave}), (\ref{eq:Veff}) the integration constant $\ls < 0$ is chosen and the restricted case ${f^b}^\lambda = 0$ is considered.

In these assumptions, a particualar exact solution of Eqs.~(\ref{eq:Einstein-traceless}), (\ref{eq:sigma-wave})  reads \cite{Pirogov2025}:
\begin{equation}
	A(r) = \left( 1 - \frac{r_g}{r}\right) \left(\frac{r}{r_h}\right)^{\frac{4\Upsilon^2}{1+2\Upsilon^2}}, \,\, %\label{eq:DH-A}\\
	 C(r) = \left(\frac{r}{r_c}\right)^{\frac{4\Upsilon^2}{1+2\Upsilon^2}}    \label{eq:DH-A-C} 
\end{equation}
\begin{equation}
	\sigma(r) = \frac{2}{1+2\Y^2} \log(\frac{r}{r_h}) \, , \label{eq:DH-sigma}
\end{equation}
where 
$r_g$ is the Schwarzschild radius of the central BH,
the scalar profile parameter
$r_h = \Upsilon \left[ - {2}/{(1+2\Upsilon^2)}\, ({1}/{\lambda_s}) \right]^{1/2}$,
and $r_c$ is fixed by the gauge-invariant requirement that at $r \to \infty$ an increase in the radial distance $\delta r$ results in  $2\pi\delta r$ increase in the measured circumference:
$r_c = r_h \left[ 1 + {2\Upsilon^2}\right]^{{1}/{2\Upsilon^2}}$
($r_c \simeq r_h e$ at  $\Upsilon^2 \ll 1$). 
In $\Upsilon \to 0$ limit the metric converges to the Schwarzschild solution.

At $\Upsilon \ne 0$ the metric (\ref{eq:DH-A-C}) features a non-flat asymptotic resulting in a kind of a gravitational confinement.
Concequently, an apparent rotation velocity for a circular orbit with the radius $r$ tends to a constant at  $r \to \infty$, 
irrespective of the central mass $M =  r_g/2G_N$ \cite{Pirogov2025}:
\begin{equation} \label{eq:v}
 v_{rot}(r) = \left[ \left( 1 + {2\Upsilon^2} \right) \, \frac{1}{2} \frac{r_g}{r-r_g} %
						  {\bf \, + \,  {2\Upsilon^2} } \right]^{1/2} \, .
\end{equation}

\paragraph*{Indications.}
This vacuum solution with an asymptotically flat rotation curve (RC) may be qualitatively used as a base for describing anomalous RCs in galaxies~\cite{flat},
though actually the flat asymptotic regime is not realized~\cite{flat-review,flat-criticism}.
More justifiably, the solution may be applied to DM dominated dwarf spheroidal galaxies (DSph) \cite{Simon:2019nxf}.
Orbital velocities of stars in outer regions of DSphs are of the order of $\sim{}10$--$30$~km/s, which 
implies $\Upsilon \sim 10^{-4}$ corresponding to the energy scale of the $\sigma$ kinetic term 
$M_S = \Upsilon \MPl \sim 10^{14}$~GeV, close to but lower than the GUT scale.
Description of variety of velocity profiles observed in DSphs (see, e.g., \cite{Hayashi:2022wnw})
whould require to study an influence of the ordinary matter on vacuum solutions, an aspect neglected so far.
Additional DM components can be considered as well.
Correlation between the ordinary (baryonic) matter and the dark halo
may arise both through the dynamical DE term (as ${\bar{\ae}}^\mu_\nu$ depends on the metric which in turn depends on the ordinary matter distribution),
as well as due to the scalar graviton coupling to possible additional DM and, through extra terms accounting  for  radiative corrections, to the ordinary matter.
Prospects of building a more realistic model of the scalar graviton DM halo were discussed in Ref.~\cite{Pirogov2025-1}.

\vspace*{1ex}

\section{Conclusion: problems and prospects}
The SBR 
presents a perspective route for future deeper merging of gravity with matter in the common framework of a spontaneous symmetry breaking.
To prove that SBR is theoretically quite consistent for merging of gravity with DE and DM,
further studies are required.
Adequacy of SBR requires a more thorough theoretical and phenomenological studies of the following problems:
finite vacuum solutions (dark holes) in the SBR framework -- static spherically symmetric and rotating ones
as a generalization of GR Schwarzschild and Kerr solutions, respectively;
connection with the conventional (SM) matter, 
coupling of the scalar graviton with matter (directly with DM, and radiatively with the ordinary matter), 
the respective modification of vacuum solutions;
anomaly of gravity  waves (tensor and scalar);
account for DE and DM sources of the scalar graviton in global solutions.

\end{document}